\documentclass[aps,prl,unsortedaddress,notitlepage,noshowkeys,nofootinbib,tightenlines,preprintnumbers,twocolumn]{revtex4-1}
\usepackage{amsmath,amssymb,amsfonts,mathrsfs,mathtools}
\usepackage{graphicx}
\usepackage{color}
\usepackage{dsfont}
\usepackage[pdftex]{hyperref}
\hypersetup{colorlinks=true, linkcolor=darkred, citecolor=blue, linktoc=page}
\definecolor{darkred}{rgb}{0.8,0.1,0.1}
\def\la{{\lambda}}
\def\La{{\Lambda}}

\def\ra{\rightarrow}

\newcommand{\be}{\begin{equation}}
\newcommand{\ee}{\end{equation}}
\def\beq#1\eeq{\begin{align}#1\end{align}}
\newcommand{\nn}{\nonumber}

\begin{document}
%\title{On the construction of de Sitter solutions in classical supergravity with orientifolds}
\title{Towards an explicit construction of de Sitter solutions in classical supergravity}
\author{Nakwoo Kim}
\email{nkim@khu.ac.kr}
\affiliation{Department of Physics and Research Institute of Basic Science,
	Kyung Hee University, 26 Kyungheedae-ro, Dongdaemun-gu, Seoul 02447, Republic of Korea}
%\affiliation{School of Physics, Korea Institute for Advanced Study, 85 Hoegi-ro, Dongdaemun-gu, Seoul 02445, Republic of Korea}
%\keywords{}
%\date{\today}

\begin{abstract}
We revisit the stringy construction of four-dimensional de-Sitter solutions using orientifolds O$8_{\pm}$, proposed by C\'ordova et al. (2019) \cite{Cordova:2018dbb}. While the original analysis of the supergravity equations is largely numerical, we obtain semi-analytic solutions by treating the curvature as a perturbative parameter. At each order we verify that the (permissive) boundary conditions at the orientifolds are satisfied. To illustrate the advantage of our result, we calculate the four-dimensional Newton constant as a function of the cosmological constant. We also discuss how the discontinuities at O$8_-$ can be accounted for in terms of corrections to the worldvolume action.
\end{abstract}

\maketitle
%\newpage
%\section{\label{sec:1}Introduction}
{\it Introduction.}
The apparent accelerating expansion of our universe is most simply explained with a positive
cosmological constant, so whether String/M-theory in lower energy description can allow it or not is a very important question. Answering it turns out to be a tough task: various
no-go theorems are established \cite{Gibbons:1984kp,deWit:1986mwo,Maldacena:2000mw}, and the constructions proposed so far are usually either not completely explicit or subject to assumptions whose validity is yet to be tested rigorously. It is even conjectured recently that de-Sitter (dS) vacua are generally not compatible in any theory of quantum gravity \cite{Obied:2018sgi}. The literature on this topic is vast, and for a review see {\it e.g.} \cite{Danielsson:2018ztv,Andriot:2019wrs}.

In this article we study a recent proposal for dS solutions in massive IIA supergravity \cite{Cordova:2018dbb}, and provide analytic results by employing a perturbative prescription. The virtue of the construction \cite{Cordova:2018dbb} is in its simplicity. 
It is done in ten dimensions, and without {\it e.g.} intersecting branes, the supergravity field equations are reduced to ordinary differential equations. The recipe is quite minimal, and one just puts orientifold 8-planes (both O$8_{+}$ and O$8_-$) in order to evade the no-go theorem \cite{Maldacena:2000mw}. Of course the solutions are non-supersymmetric, so the stability is not guaranteed. They also suffer from singularities at the orientifolds, but otherwise we are given a relatively straightforward, well-defined mathematical problem of analyzing coupled nonlinear differential equations with delta-function sources representing the O8-planes.

The current work is also strongly motivated by the criticism in \cite{Cribiori:2019clo}, which came up with a no-go argument, according to which the numerical solutions in \cite{Cordova:2018dbb} are invalidated unless extra ingredients {\it e.g.} O6-planes are added. In a more recent work however \cite{Cordova:2019cvf}, the authors of \cite{Cordova:2018dbb} 
%have presented a more general class of dS solutions which involve O6-planes as well as O8, and they 
have presented a refined version of boundary conditions near O$8_-$, advocating the existence of numerical solutions which satisfy such {\it permissive}, {\it i.e.} less stringent, requirements.
% in the original construction involving only O8-planes
The issue here is basically whether one should equate only the leading coefficient of two divergent quantities at the singularity, or more restrictively the sub-leading finite part as well. The permissive condition presumes that the discontinuity of the finite part will be fixed when string corrections are taken into account.

In our computation we verify that while the permissive boundary conditions are satisfied, the {\it restrictive} ones are not satisfied just as the authors of \cite{Cribiori:2019clo} pointed out.
Using our result, any physical quantity can be calculated as a series in $\La$, the cosmological constant. As an example we calculate the four-dimensional Newton constant. We also construct extra boundary terms at O$8_-$, with which the solutions do respect the restrictive boundary conditions.

%Our plan is as follows. Sec.\ref{sec:2} summarizes the technical setup of \cite{Cordova:2018dbb}, giving the differential equations we need to solve, and the two versions of boundary conditions near O8${}_-$. Sec.\ref{sec:3} is the main part where we introduce and execute our perturbative prescription up to $\La^{20}$. We conclude in Sec.\ref{sec:4}.

%%%%%%%%%%%%%%%%%%%%%%%%%%%%%%
%\section{\label{sec:2}Supergravity equations and boundary conditions}
{\it The setup and the boundary conditions.}
The proposal in \cite{Cordova:2018dbb} is to consider massive IIA supergravity, and add O8-planes. More concretely, one employs the following metric ansatz in string frame,
\be
ds^2 = e^{2W} ds^2_{{\rm dS}_4} + e^{-2W} \left( dz^2 + e^{2\lambda} ds^2_{M_5} \right) . 
\ee
Namely, the ten-dimensional spacetime comprises the dS${}_4$ spacetime with warp factor $e^{2W}$, a compact direction parametrized by $z$, and a negatively-curved Einstein manifold $M_5$. We have three functions - $W(z),\lambda(z)$ and the dilaton $\phi(z)$ - to be determined.

In order to evade the no-go theorem for dS vacua in supergravity through dimensional reduction \cite{Gibbons:1984kp,deWit:1986mwo,Maldacena:2000mw}, one allows a negative-tension object at $z=z_0$ (O$8_-$), in addition to an O$8_+$ plane at $z=0$, where $z$ is periodic as $z\sim z+2z_0$. The orientifolds in supergravity are treated as a delta-function-like source of tension and charge, and their full backreaction will be considered. The field equations are then reduced to \cite{Cordova:2018dbb}
\beq
\MoveEqLeft W'' +W'(5\lambda-2\phi)'  - \Lambda e^{-4W} 
\nn\\
&- \frac{1}{4} F^2_0 e^{2(\phi-W)}= \frac{1}{\pi} e^{\phi-W} \sigma,
\label{eq1}
\\
\MoveEqLeft (W+2\phi-5\lambda)''+W'(5\la+2\phi)'-8(W')^2 - 5 (\la')^2 
\nn\\
&+ \frac{1}{4} F^2_0 e^{2(\phi-W)} = \frac{1}{\pi} e^{\phi-W}\sigma,
\\
\MoveEqLeft (W-\la)'' + (W-\la)'(5\la-2\phi)'  
\nn\\
&-\frac{4\La}{5} e^{-2\la} + \frac{1}{4} F^2_0 e^{2(\phi-W)} = -\frac{1}{\pi} e^{\phi-W}\sigma,
\label{eq3}
\\
\MoveEqLeft 4(W')^2 - 10 (\la')^2 - 2 (\phi')^2 + 2\phi' (5\la' - W')
\nn\\
&  +2 \La e^{-4W}   -2\La e^{-2\la} - \frac{1}{4} F^2_0 e^{2(\phi-W)} = 0 .
\label{eq4}
\eeq
Here $F_0=-2/\pi$ is the mass parameter of massive IIA, $\Lambda$ is the cosmological constant of dS${}_4$, and 
%$\kappa$ is defined so that 
the Ricci scalar of $M_5$ is $5\kappa$. Without losing generality, we set $\kappa=-4\Lambda/5$ for convenience. 
The orientifolds manifest themselves as the delta-function source $\sigma := \delta (z) - \delta (z-z_0)$. 

One can verify that the above equations \eqref{eq1}-\eqref{eq3} can be derived from the following 
effective action.
\begin{align}
     S_{\rm eff} & = \int^{z_0}_0 dz \, e^{5\lambda-2\phi}
     \Big[ 4(W')^2 - 10 (\la')^2 - 2 (\phi')^2 
     \nn\\
     & + 2\phi' (5\la' - W') 
      +{2} \La (e^{-2\la}  -  e^{-4W} )
     \nn\\
     & + \frac{1}{4} F^2_0 e^{2(\phi-W)} +  \frac{2}{\pi} e^{\phi -W} ( \delta(z) - \delta(z-z_0))\Big]
     %\Big] 
     ,
     \label{effaction}
\end{align}
and the zero-energy Hamiltonian constraint \eqref{eq4}.
%for three functions $W,\la,\phi$ 
%(taken verbatim from \cite{Cordova:2019cvf}):
%\begin{widetext}
%\begin{small}
%\beq
%\label{eq1}
%\MoveEqLeft W'' +W'(5\lambda'-2\phi')  - \Lambda e^{-4W} \nn\\ & 
%- \frac{1}{4} F^2_0 e^{2(\phi-W)}
%\nn\\
%& 
%= \frac{1}{\pi} e^{\phi-W} \sigma,\\
%\MoveEqLeft (W+2\phi-5\lambda)''+W'(5\la+2\phi)'-8(W')^2 - 5 (\la')^2 \nn\\
%& + \frac{1}{4} F^2_0 e^{2(\phi-W)} = \frac{1}{\pi} e^{\phi-W}\sigma,\\
%\MoveEqLeft (W-\la)'' + (W-\la)'(5\la-2\phi)' + \kappa e^{-2\la} \nn\\
%& + \frac{1}{4} F^2_0 e^{2(\phi-W)} = -\frac{1}{\pi} e^{\phi-W}\sigma,
%\label{eq3}\\
%\MoveEqLeft 4(W')^2 - 10 (\la')^2 - 2 (\phi')^2 + 2\phi' (5\la - W)' +2 e^{-4W} \Lambda 
%\nn\\
%& +\frac{5}{2} \kappa e^{-2\la} 
% - \frac{1}{4} F^2_0 e^{2(\phi-W)} = 0 ,
%\label{eq4}
%\eeq
%\end{small}
%\end{widetext}
%where prime denotes $(\bullet)':=d(\bullet)/dz$, 

The usual prescription for 2nd-order differential equations with a delta-function source is that the functions themselves are continuous while the first derivatives exhibit discontinuity. The subtlety here is that the functions $W,\la,\phi$ are divergent near the negative-tension object O8${}_-$ at $z=z_0$. On the other hand, at $z=0$ the functions $W,\la,\phi$ are finite and $\sigma(z)$ can be treated in the standard way. 
\be
\label{bc0}
\lim_{z\ra 0^+} e^{W-\phi}f_i' =  -({4F_0})^{-1} = ({2\pi})^{-1},
\ee
where $f_i\equiv \left\{ W, \phi/5, \la/2 \right\}$ collectively represent the functions to solve for.
The functions $f_i,f_i'$ are all finite at $z=0$, so there is no subtlety with \eqref{bc0}.

Now let us do the same with 
the boundary condition at $z=z_0$. 
From the equations of motion, one might naively want to impose 
\be
\label{restrict}
\lim_{z\ra z_0^-} \left(f_i'- ({2\pi})^{-1} e^{\phi-W}\right)=0 \quad {(restrictive)}.
\ee
But in fact it is too restrictive, since it equates not only the leading divergent part but also the sub-leading and finite part. It was thus proposed \cite{Cordova:2019cvf} that one should impose the following condition which in fact fixes only the leading logarithmically divergent part,
\be
\label{permi}
\lim_{z\ra z_0^-} e^{W-\phi}f_i' = ({2\pi})^{-1} \quad {(permissive)}.
\ee
This prescription is supported by the observation that 
a family of successfully tested AdS/CFT duals involving orientifolds exhibit a curvature singularity with 
the same property \cite{Apruzzi:2013yva,Apruzzi:2017nck}. From a more technical viewpoint, the  permissive boundary condition is obtained when the field variations are restricted to $L^2$ space, while the restrictive one is derived when the field variations are required to be smooth \cite{Cordova:2019cvf}.

%%%%%%%%%%%%%%%%%%%%%%%%%%%%%%%%
%\section{\label{sec:3}Perturbative Solutions}
{\it Perturbative Solutions.}
Our idea is to solve \eqref{eq1}-\eqref{eq4} perturbatively. We will start with the case when $\Lambda=\kappa=0$ in the above, and treat the remaining terms in question as perturbation. To expedite our analysis let us introduce ($H_i=e^{-4f_i}$ in the notation of \cite{Cordova:2019cvf})
\beq
\{ w , p ,q \} &:= \{ e^{-4W}, e^{-4\phi/5}, e^{-2\la} \}.
\eeq
%Note that $f_i=-\tfrac{1}{4}\{ \log w , \log p , \log q \}$.

One can then easily check, if we choose to put O$8{}_+$ and O$8{}_-$ at $z=0$ and $z=1$ respectively,
\beq
\label{zs}
\{ w_0 , p_0 ,q_0 \} 
:=\left\{ \frac{\pi^4c_1^5}{16} ,
 c_1,
\frac{\pi^4c_3}{16} 
\right\}|1-z|,
\eeq
satisfy the equations, where $c_1,c_3$ are constants. We note that the boundary condition at $z=0$ is satisfied, and the behaviour at $z=1$ implies that the quantities in the restrictive boundary condition diverge but match exactly, while the permissive one is  satisfied as an equality between finite quantities. 
%From now on we treat $z$ defining an interval, $0<z<1$, with O8-planes 

From now on let us assume that $z$ lies in the interval $0<z<1$, and the functions satisfy 
appropriate limiting behaviour at $z=0$ and $1$, as dictated by the boundary conditions. Our strategy is to solve the equations for non-zero $\Lambda$, by substituting 
\beq
w(z) &= w_0(z)(1  + \sum_{n=1} \pi^{4n} \Lambda^n  w_n(z)) ,
\\
p(z) &= p_0(z)(1  + \sum_{n=1} \pi^{4n} \Lambda^n  p_n(z)) ,
\\
q(z) &= q_0(z)(1  + \sum_{n=1} \pi^{4n} \Lambda^n  q_n(z)) ,
\eeq
into the equations of motion. Organising them as a power series in $\La$, we obtain linearized differential equations for $w_n,p_n,q_n$, which we can solve exactly. One then demands that the (permissive) boundary conditions be satisfied at both $z=0$ and $z=1$. Then the result is straightforwardly extended to $-1<z<0$ since the functions are all even, and periodic with 
$z\sim z+2$.

Let us comment that this approach is reminiscent of recent works \cite{Kim:2019feb,Kim:2019rwd,Kim:2019ewv,Kim:2020unz}, where supergravity solutions in various holographic contexts are constructed using a perturbative prescription. A notable difference here is that we are looking for non-supersymmetric solutions, so instead of first-order BPS relations we have second-order differential equations, and the analysis is more challenging.

The equations for $w_1,p_1,q_1$ are given as follows, where the source terms are omitted and will be taken care of by imposing the permissive boundary condition.
%\begin{widetext}
\beq
\MoveEqLeft 4(1-z)^2 w_1'' -10(1-z)(p_1'-q_1') 
\nn\\&
+2 w_1 - 10 p_1 + c_1^5(1-z)^3 = 0 ,
\label{lin1}
\\
\MoveEqLeft 2(1-z)^2(w_1''+10p_1''-10q_1'')  
\nn\\&
+(1-z) ( 2w_1' + 5 p_1' - 15 q_1')-w_1 + 5 p_1 = 0,
\label{lin2}
\\
\MoveEqLeft 10(1-z)^2(w_1''-2 q_1'') +25 (1-z) (p_1'-q_1') 
\nn\\&
- 5 w_1 +25 p_1 +2 c_3 (1-z)^3 = 0 ,
\\
\MoveEqLeft (1-z) (w_1' + 5 p_1' - 10 q_1' ) - w_1 + 5 p_1 
\nn\\&
+ ( c_1^5 -c_3 ) (1-z)^3 = 0 .
\label{lin4}
\eeq 
%\end{widetext}
%\begin{widetext}
%\beq
%& w_1'' -\frac{5(p_1'-q_1')}{2(1-z)}
%\nn\\&
%+\frac{ w_1 - 5 p_1}{2(1-z)^2} =-\frac{c_1^5(1-z)}{4} ,
%\label{lin1}
%\\
%& w_1''+10p_1''-10q_1'' +\frac{2w_1' + 5 p_1' - 15 q_1'}{2(1-z)}
%\nn\\&
%=\frac{w_1 - 5 p_1}{2(1-z)^2},
%\label{lin2}
%\\
%& w_1''-2 q_1'' +\frac{5 (p_1'-q_1')}{2(1-z)}
%\nn\\&
%- \frac{w_1 -5 p_1}{2(1-z)^2} =- \frac{c_3 (1-z)}{5} ,
%\\
%& w_1' + 5 p_1' - 10 q_1' 
%\nn\\&
%- \frac{w_1 - 5 p_1}{1-z} = ( c_3-c_1^5 ) (1-z)^2  .
%\label{lin4}
%\eeq 
%\end{widetext}

One can find the general solutions explicitly, with five integration constants. It is indeed the case that the restrictive boundary conditions are too strong and no choice of the integration constants can satisfy them. On the other hand, permissive boundary conditions and the requirement to maintain the position of O$8_-$ at $z=1$, by setting $w_1=p_1=q_1=0$ at $z=1$, completely fix the solution, with an extra relation $c_3=c^5_1$.

We can explicitly see what goes wrong with the restrictive boundary conditions. Near $z=1$, our 
${\cal O}(\La)$ result gives 
\beq
4f_1' &= -\frac{w'}{w} = \frac{1}{1-z}+\frac{\pi ^4c_1^5}{128}
%4f_1' &= -{w'}/{w} = \tfrac{1}{1-z}+\tfrac{\pi ^4c_1^5}{128}
   \left(1 +34 z -17  z^2 \right)
   \Lambda ,
%   +O\left(\Lambda^2\right)
   \\
4f_2' &=  -\frac{p'}{p} =\frac{1}{1-z}+\frac{\pi ^4c_1^5}{128} \left(1+2
    z - z^2 \right) \Lambda ,
%   +O\left(\Lambda^2\right)
   \\
4f_3' &= -\frac{q'}{q} = \frac{1}{1-z}+\frac{\pi ^4c_1^5}{
   640} \left(5+26  z  -13  z^2 
    \right) \Lambda .
%    +O\left(\Lambda ^2\right)
\eeq
Obviously, because the ${\cal O}(\La)$ parts of $f_i'$ here all take distinct values at $z=1$, the  restrictive condition is violated. 
%What happens is that the permissive condition only demands the leading $(1-z)^{-1}$ terms to be the same, which is already strong enough to completely fix $w_1,p_1,q_1$ as above. 
%Obviously in the above ${\cal O}(\La)$ terms are all distinct, which is responsible for the discontinuity of {\it e.g.} $W'-\phi'/5$ at $z=z_0$, as pointed out in \cite{Cribiori:2019clo}.

At higher orders of $\Lambda$, one proceeds essentially in the same way. The homogeneous part of the equations for $w_n,p_n,q_n$ are the same as $n=1$, while the inhomogeneous part is determined by the solutions for small $n$ and gets complicated gradually. One also needs to allow $\Lambda$-dependence in  the relation between $c_3$ and $c_1$. Namely,
\be
c_3(c_1;\La) = c_1^5 + \sum_{n=1} \pi^{4n} \Lambda^n {\mathfrak c}_{n+1}.
\ee
and ${\mathfrak c}_{n}$ can be determined uniquely as well.
%and the result is like
%\begin{enumerate}
%{\mathfrak c}_2 = \tfrac{3 }{40}c_1^{10} , \quad {\mathfrak c}_3 = \tfrac{9 }{1024}c_1^{15}, \quad
%{\rm etc} . 
%\ee

We have done the iterative computations up to $\La^{20}$ explicitly, although we present only the results up to $\Lambda^2$ below. 
%It turns out that, also at higher orders of $\La$, the functions $w,p,q$ always take a polynomial form in $z$.
\begin{widetext}
\beq
w(z) &= -\tfrac{\pi ^4c_1^5 }{16}  (z-1)
-\tfrac{\pi ^8 c_1^{10} }{6144}(z-1)^2
   \left(17 z^2-34
   z-37\right)\La
   \nn\\
&   -\tfrac{\pi ^{12} c_1^{15} }{103219200}(z-1)^2
   \left(10117 z^5-50585 z^4+30547
   z^3+95579 z^2-68464
   z-77107\right)\La^2,
   \\
 %  &
 %  -\frac{\pi ^{16} c_1^{20} (z-1)^2 \left(6598273 z^8-52786184
 %  z^7+113458549 z^6+36171442 z^5-324770015 z^4+122742328
 %  z^3+355949671 z^2-184706786
 %  z-211504643\right)}{1981808640000}
 %  \\
p(z) &= -c_1 (z-1)
-\tfrac{\pi ^4c_1^6 }{384}   (z-1)^2 \left(z^2-2 z-5\right)\La
   \nn\\&   
   -\tfrac{\pi^8 c_1^{11}}{2150400} (z-1)^2
   \left(199 z^5-995 z^4-551 z^3+593
   z^2-888 z-2369\right)\La^2,
   \\
q(z) & = -\tfrac{\pi^4c_1^5 }{16}  (z-1)
-\tfrac{\pi ^8 c_1^{10} }{30720}(z-1) \left(13 z^3-39 z^2-15
   z+185\right) \La
   \nn\\
   &
   -\tfrac{\pi^{12} c_1^{15} }{516096000}(z-1) \left(9257 z^6-55542 z^5+54120
   z^4+54260 z^3-120915 z^2-43215 z+385535\right)\La^2.
\eeq
\end{widetext}
%Recall that in the above the domain is $0<z<1$, and in order to extend it we exploit reflection symmetry $z\ra -z$ and periodicity $z\sim z+2$. 
%From the above, one naturally expects that the coefficient of $\La^n$ in $w,p,q$ should be a $(3n+1)$-th order polynomial, and it turns out to be the case. 

As an example of what one can do using our result,
we evaluate the supergravity action, and read off the Newton constant from the coefficient of the curvature scalar.
%\beq
%S &= \frac{1}{(2\pi)^7 l^8_s}\int d^{10}x \sqrt{g_{10}} \left[ R_{10} + \cdots \right]
%\nn\\
%&= \frac{1}{16\pi G_4} \int d^{4}x \sqrt{g_{4}} \left[ R_{4} +\cdots \right]
%\eeq
%Considering the dimensional reduction from ten down to four-dimensions, we have
\be
M^2_P = 2\kappa^2_{10} {\rm vol}({M_5}) \int^1_0 e^{-4W-2\phi+5\la} dz ,
\ee
where $M_P$ is the four-dimensional Planck mass, and $\kappa_{10}$ is the ten-dimensional gravitational constant. The integral at hand is
\beq
\label{newton}
%\int^1_0 e^{-4W-2\phi+5\la} dz &=
%& s(\tilde\La)\equiv
& 
\int^1_0 w \left( p/q \right )^{5/2} dz 
= \tfrac{1}{\pi^6 c_1^5}\Big( 32 -{41} {\tilde \La} 
-\tfrac{4297 }{896} {\tilde \La}^2 \nn\\
&-\tfrac{5890851 }{788480} {\tilde \La}^3
-\tfrac{13579752323 }{1252392960} {\tilde \La}^4
 -\tfrac{7590791178245599}{449659168358400} {\tilde \La}^5
\nn\\
& -\tfrac{227867772628905066871}{8201783230857216000} {\tilde \La}^6
%\nn\\
 -\tfrac{33203220336431649984314981}{697217188888710217728000} {\tilde \La}^7
\nn\\
& -\tfrac{1464517396860440958310400521603}{17402541034662207034490880000} {\tilde \La}^8
\nn\\
&-\tfrac{449002321297358563183375027089959}{2949049736378235571201376256000} {\tilde \La}^9
\nn\\
&-\tfrac{1332929089357763777590588348841714411807}{4749566031896881608884512987545600000} {\tilde \La}^{10}
\nn\\
%&-\tfrac{13321957768047533723033579576552293271060043373}{25366102297872313546201696202844536832000000} {\tilde \La}^{11}
%\nn\\
%&-\tfrac{3407405179885895507120617642723513687732714223853749}{3423815023757613393212120146675144203436032000000} {\tilde \La}^{12}
%\nn\\
&- \cdots - {1.09457\times 10^{5}} {\tilde \La}^{19}- {2.18919\times 10^{5}} {\tilde \La}^{20}
%&-\frac{7590791178245599 }{44965916835840000000} {\tilde \La}^5 \nn\\
%&-\frac{227867772628905066871 }{8201783230857216000000000} {\tilde \La}^6 \nn\\
%&-\frac{33203220336431649984314981 }{6972171888887102177280000000000}{\tilde \La}^7
\Big),
\eeq
where ${\tilde \La}=\pi^4c_1^5\La/10$. This function 
%$s(\tilde\La)$ 
is monotonically decreasing, and it vanishes when  ${\tilde\La}\approx 0.53$.
%\be
%\frac{1}{16\pi G_4} = \frac{2{\rm vol}(M_5)}{(2\pi)^7l^8_s} \int^1_0 e^{-4W-2\phi+5\la} dz
%\ee
%\be
%\frac{1}{16\pi G_4} = \frac{1}{(2\pi)^7l^8_s}
%\ee

\begin{figure}
\centering
\includegraphics[width=.35\textwidth]{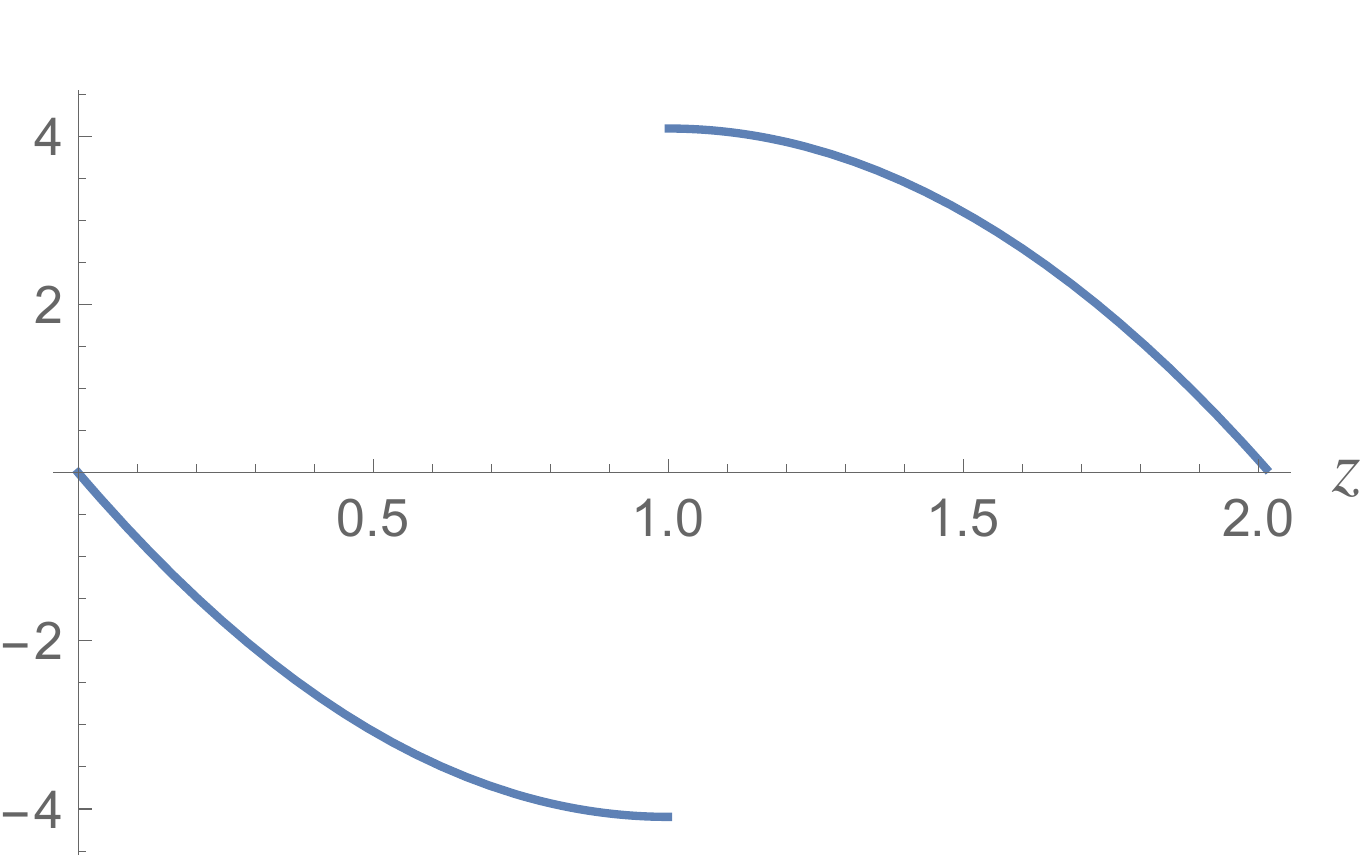} 
\caption{\label{fig1}A plot of $w^{-1}p^{7/2}{q^{-5/2}}\left(w/p\right)'$, for $\La=0.1,c_1=0.5$. The discontinuity at $z=1$ is also pointed out in \cite{Cribiori:2019clo}.}
\end{figure}

Recall that the computation of the lower-dimensional cosmological constant is exactly how one derives the no-go theorem \cite{Gibbons:1984kp,deWit:1986mwo,Maldacena:2000mw,Cribiori:2019clo}. Indeed, one can check that a particular linear combination of \eqref{eq1}-\eqref{eq4} gives
\beq
%\left[ \frac{p^{7/2}}{wq^{5/2}}\left(\frac{w}{p}\right)'\right]' + 4\La w\left( \frac{p}{q} \right )^{5/2} = 0 . 
\left[  w^{-1}p^{7/2} q^{-5/2} \left({w/p}\right)'\right]' + 4\La w\left( {p/q} \right )^{5/2} = 0 . 
\eeq
Taken at face value,
mathematical consistency would require that 
%Note that the source terms from the effective action \eqref{effaction} are here all eliminated, not ignored. Taken at face value, consistency requires that 
the function inside the square bracket should be discontinuous at $z=z_0$ (see Fig.\ref{fig1}), which calls for new delta-function source terms \cite{Cribiori:2019clo}. But this is exactly what the restrictive boundary condition demands. After all, classical supergravity is an effective theory which breaks down at O$8_-$. We adopt the permissive conditions since O$8_-$ is a legitimate object in string theory, and we expect the sub-leading discontinuity above should be also cured once we include stringy corrections. 

For the final verdict we should in principle wait until all the correction terms in the action are identified, but let us carry out a relatively simple test instead. 
Would it be possible to add certain extra boundary terms at $z=z_0$ to \eqref{effaction}, so that our explicit solutions satisfy the restrictive version of the modified boundary conditions? The answer is in the affirmative, it turns out.

As a technical assumption, 
we allow the correction terms contain the fields and the parameter $\La$ but not $c_1$, and require they make finite contribution to discontinuity of $f_i'$ at $z=z_0$. They should be compatible with string perturbation, which implies only higher orders in $e^\phi$ are allowed. Then the most general form of the correction terms should be 
\beq
 %\int^{z_0}_0 
 \Big[
 e^{5\lambda-2\phi} \sum_{n=1}^\infty \frac{\La^n}{\pi^n} e^{n(\phi-5W)} G_n (e^{4W-2\la}) 
 %\delta(z-z_0)
 \Big]_{z=z_0}
 . 
\eeq
We demand $G_n(1)=0$, since $e^{4W-2\la}=1$ when $\La=0$, and we know we do not need a correction term in that case. Let us henceforth write
\beq
G_n (s) = \sum_{k=1}^\infty g_{n,k} (1-s)^k . 
\eeq
One then studies how $G_n$ affects the equations \eqref{eq1}-\eqref{eq3}, and see if the restrictive boundary condition can be simultaneously met, by choosing $g_{n,k}$ appropriately.

We have verified that \eqref{restrict} with correction terms can be achieved indeed, but not all $g_{n,k}$ are fixed uniquely. It is interesting though that at least the first three coefficients are determined,
\beq
g_{1,1} = 2 , \quad  g_{1,2} = -\tfrac{37}{48}, \quad g_{2,1} = \tfrac{53}{50} . 
\eeq
%And we have discovered a particular scheme where all the coefficients are uniquely fixed. If we assume $g_{n,k}=0$ for $k=3,4,\cdots$, we obtain
%\beq
%g_{2,2} = -\tfrac{37451}{61600}, \quad g_{3,1} = \tfrac{1927}{1375}, \quad g_{3,2} = -\tfrac{28443881}{36960000}, 
%\eeq
%and so on. 
Other than these, there are many terms which make the same effect on \eqref{eq1}-\eqref{eq3} and our computation alone cannot distinguish them.

It is an intriguing question now whether the correction terms obtained above can be shown to arise naturally in string theory. Although giving a full answer is beyond our scope in this paper, let us point out that the terms with $g_{1,1}$ and $g_{1,2}$ may come from a boundary action of the worldvolume curvature-squared, {\it e.g.} $\int_{O8} e^{-\phi}\sqrt{g_9} (R_{9})^2$. 
%For other terms, it is not clear to us how they come from a world-volume covariant expression.
%%%%%%%%%%%%%%%%%%%%%%%
%\section{\label{sec:4}Discussion}

{\it Discussion.}
In this paper we have solved the supergravity equations for the dS$_{4}$ construction in \cite{Cordova:2018dbb}, and obtained the solution explicitly as a power series in the four-dimensional cosmological constant $\La$. Our explicit formulae, although it is unlikely we can sum them exactly, enable us to calculate physical quantities as a series expansion form in $\La$. Of course a result like \eqref{newton} should be taken with a grain of salt, because of the stringy correction terms needed to resolve the orientifold singularity. For this particular quantity however, the integrand in \eqref{newton} vanishes at $z=1$, so we expect the corrections are suppressed. Additionally, we expect one can also do the stability analysis and calculate tachyon potential \cite{Garg:2018reu}, compute the fluctuation spectrum etc. with our results.

Just like our previous works \cite{Kim:2019feb,Kim:2019rwd,Kim:2019ewv,Kim:2020unz}, the result here
lends further support to the perturbative prescription as a powerful alternative to numerical analyses of supergravity equations which are generically nonlinear. We comment that an important requirement for our prescription is an explicit, and preferably simple, unperturbed solution, like \eqref{zs}.
It is just the D8-brane solution with flat world-volume, as one can easily see. We expect there are many other systems to which we can apply a similar method, 
and the dS${}_4$ construction using O$8_+$--O$6_-$ in \cite{Cordova:2019cvf} is one of them which we hope to address in a future work.

%in particular AdS or dS constructions involving orientifolds \cite{Cordova:2018eba,Cordova:2019cvf,Marchesano:2020qvg}.
%\begin{acknowledgments}
{\it Acknowledgments.}
We thank D. Junghans for comments and encouraging us to calculate the corrections to the O8 action using our results.
This work was supported by the National Research Foundation (NRF) grant 2019R1A2C2004880.
%\end{acknowledgments}

\bibliography{ds4}

\end{document}